\documentclass[aps,prl,groupedaddress,twocolumn,10pt]{revtex4}
\usepackage{amsmath}%
\usepackage{amsfonts}%
\usepackage{amssymb}%
\usepackage{graphicx}
\usepackage{braket} 
\usepackage{bbold}
\usepackage{textcomp}
\usepackage{xspace}
\usepackage{color}
\newcommand{\vect}[1]{{\bf{#1}}}

\newcommand{\uvect}[1]{{\bf{\hat{#1}}}}
\newcommand{\C}{\textsuperscript{13}C\xspace}
\newcommand{\Cs}{\textsuperscript{13}C spins\xspace}
\newcommand{\eq}[1]{{Eq.~(\ref{#1})}}

\newcommand{\ccite}[1]{\,\cite{#1}}

\begin{document}

%Title of paper
\title{Sensing distant nuclear spins with a single electron spin}

% Place the author information here.  Please hand-code the contact
% information and notecalls; do *not* use \footnote commands.  Let the
% author contact information appear immediately below the author names
% as shown.  We would also prefer that you don't change the type-size
% settings shown here.

\author{Shimon Kolkowitz,$^{\dagger}$
Quirin P. Unterreithmeier,$^{\dagger \ast}$ 
Steven D. Bennett \&
Mikhail D. Lukin}
\affiliation{Department of Physics, Harvard University, Cambridge, MA 02138, USA}

\begin{abstract}
We experimentally demonstrate the use of 
a single electronic spin to measure the quantum dynamics of distant
individual nuclear spins from within a surrounding spin bath.
Our technique exploits coherent control of the electron spin, allowing us to
isolate and monitor nuclear spins weakly coupled to the electron
spin.
Specifically, we detect the evolution of distant individual \C nuclear spins coupled to single nitrogen vacancy centers in a diamond lattice with hyperfine couplings down to a factor of 8 below 
the electronic spin bare dephasing rate.
Potential applications to nanoscale magnetic 
resonance imaging and quantum information processing are discussed. 
\end{abstract}
\date{\today}

\maketitle

Detection and control of the magnetic signals from nuclear spins is an important problem in science and technology.   
Single nuclear spin detection remains an outstanding goal in 
magnetic resonance imaging (MRI), and could have far-reaching implications
from physics to medicine\ccite{Rugar:2004wi,Degen:2009wa,Hammel2007}.
Likewise, nuclear spins stand out in quantum information science for their exceptional isolation from their environment, making them attractive qubit candidates. However, as this isolation results from their weak magnetic moments, individual nuclear spins are extremely difficult to detect and control. Utilizing the electronic spin associated with a single nitrogen-vacancy (NV) center in diamond to both detect and control surrounding nuclei is a promising approach to this challenge, and significant efforts are currently directed towards this goal. Early work demonstrated that NV centers can be used to sense strongly coupled proximal nuclear spins \ccite{Jelezko2004,Childress2006, Hanson2008}.  Subsequently, this approach has been used to create a few-qubit quantum register \ccite{Dutt2007, Jiang2009}, perform simple quantum algorithms \ccite{Jelezko2004, Sar:2012fk}, implement single shot readout of both nuclear and electronic spins \ccite{Neumann2010, Childress:2011wv}, and demonstrate a multi-second quantum memory at room temperature using a single coupled \C nuclear spin in isotopically purified diamond\ccite{Maurer2012}. 

In this Letter we show that the NV electronic spin can be used to isolate and probe the quantum dynamics of distant, weakly coupled individual Carbon-13 nuclear spins.  Our approach relies on dynamical decoupling sequences, which enhance the sensitivity to individual nuclear spins while suppressing NV electronic spin decoherence\ccite{Taylor:2008ub,deLange:2010ue, Hall:2010vs, Naydenov:2011uz, deLange:2011wm,Ryan2010,Dobrovitski2012}.  Crucially, this allows us to observe the coherent evolution of nuclear spins whose coupling to the NV is weaker than the limit imposed by the NV spin's inhomogenous dephasing rate, $1/T_2^*$.  
We demonstrate this capability by identifying and measuring the coupling to single nuclear spins 
from amongst a bath of naturally abundant \C nuclear spins. The ability to isolate spins from within a bath is important for applications where surrounding nuclear spins cannot be avoided; moreover, it enables the use of these nuclear spins as a resource. In particular, our technique could be extended to harness weakly coupled nuclear spins in quantum registers\ccite{Jelezko2004,Jiang2009,Neumann2010, Childress:2011wv,Maurer2012,Sar:2012fk}, to investigate the spatial extent of the NV electronic wave function\ccite{Gali2008, Gali2009, Smeltzer2011, Jacques2012}, or to detect individual spins outside of the diamond lattice for single spin MRI applications \ccite{Taylor:2008ub,Maze:2008ws, Rugar:2004wi, Degen:2009wa}.  

 \begin{figure}[b]
\begin{center}
\includegraphics{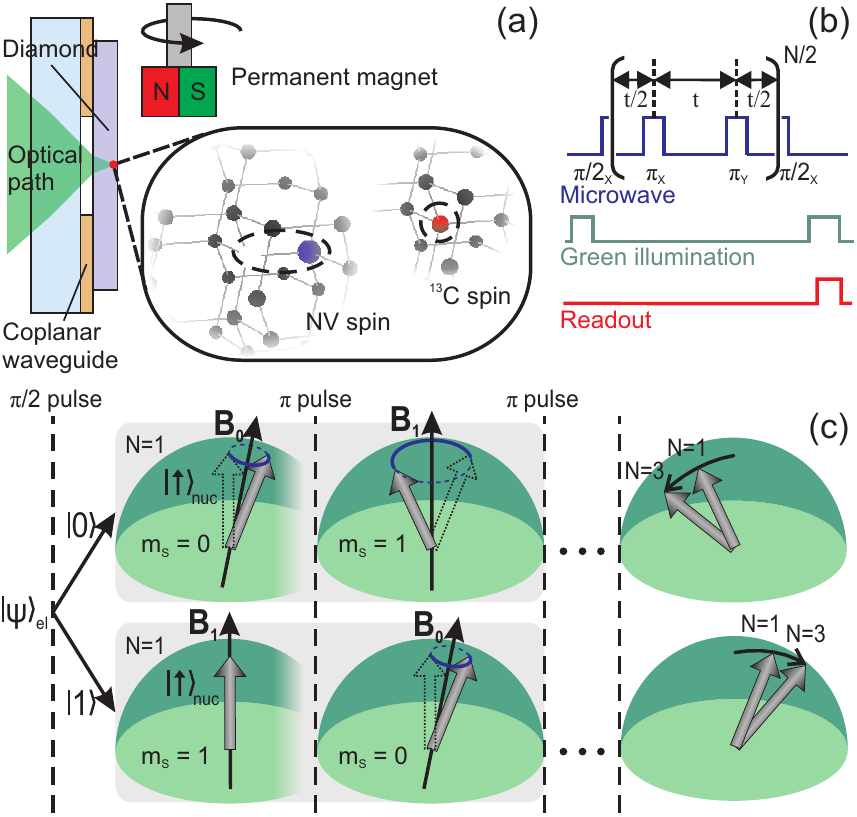}%
\end{center}
 \caption{(a) Schematic setup.  We measure
 shallow-implanted NV centers weakly coupled to 
  \C nuclear spins.  A dc magnetic field is applied 
  using a movable permanent magnet. 
  (b) Pulsed spin manipulation and readout. 
 (c) Central concept.  Conditional evolution of the \C on the Bloch sphere during NV spin manipulation, for \C spin initially in state $\ket{\uparrow}$. 
Upper (lower) panel shows \C evolution with the NV spin initially
in state $\ket 0$  ($\ket 1$). 
Gray boxes show evolution during spin echo (single $\pi$-pulse). 
Additional $\pi$-pulses push
conditional nuclear spin evolution further apart,
increasing its entanglement with the NV and thus
the signal. For clarity a sequence with an odd number of $\pi$-pulses is shown; the result is qualitatively the same for the even numbered sequences used in this work, but is more difficult to visualize.}
\label{fig_1}%
\end{figure}

The central idea of this work is
depicted in Fig.\,\ref{fig_1}c, and can be understood in
terms of the coherent evolution of a single \C nuclear spin
interacting with the NV electronic spin sensor.
Through their interaction, the magnitude and orientation of the
local magnetic field experienced
by the \C spin depends on the NV spin state.
As a result, when the NV spin is prepared in a superposition of its energy eigenstates, the \C spin undergoes conditional Larmor precession around two different axes. 
During this precession the electron and nuclear spin become entangled and disentangled, resulting in collapses and revivals in NV spin coherence \ccite{Childress2006}.
However, if the NV-nuclear coupling is weak then
the two precession axes are similar and the resulting entanglement is small.
The key idea of this work is to increase the degree of entanglement, and thus the measurable signal, by applying periodic $\pi$-pulses, flipping the NV spin with a frequency matched to the precession of the \C spin.  
As shown schematically in Fig.\,\ref{fig_1}c, this constructively enhances the 
conditional evolution of the \C state further apart, increasing its 
entanglement with the NV spin.

To describe the experimental sequence shown in Fig.~\ref{fig_1}b, we consider the mutual interaction between a single NV electronic spin-$1$ and a single \C nuclear spin-$\frac{1}{2}$, which is governed by the Hamiltonian ($\hbar = 1$)
\begin{equation}
	H = \Delta S_z^2 - \mu_e B_z S_z + (\mu_n\vect B + S_z \vect A) \cdot \vect I .
%	+ S_z \vect A \cdot \vect I.
\label{eq_hamiltonian}
\end{equation}
Here, $\vect S$ ($\vect I$) and $\mu_e$ ($\mu_n$) are the electronic (nuclear) spin and magnetic moment respectively, 
$\vect B$ is the external magnetic field, and $\vect A$ is the hyperfine interaction. 
Due to the large zero field splitting $\Delta/2\pi \simeq 2.87$\,GHz, we have
made the secular approximation conserving $S_z$. 
The
nuclear spin evolves conditionally on the NV spin state according to 
$H_{{\rm nuc}}[m_s] = \frac{\omega_{m_s}}{2} \boldsymbol \sigma \cdot {\bf n}_{m_s},$
where $\boldsymbol \sigma$ is the vector of Pauli spin matrices. Thus
the nuclear spin precesses about an effective magnetic
field axis ${\bf n}_{m_s} = ( \mu_n{\bf B} + m_s {\bf A}) / \omega_{m_s}$ 
with Larmor frequency $\omega_{m_s} =  \left| \mu_n{\bf B} + m_s {\bf A}  \right|$
when the NV spin is in state $\ket{m_s}$.
We initialize the NV spin in a superposition of
$\ket 0$ and $\ket 1$ and apply a periodic series of $N ~\pi$ pulses spaced by evolution time $t$, during which $\ket 0$ and $\ket 1$ accumulate a relative phase. 
Converting the phase into a population difference, the normalized fluorescence signal at the end of the
measurement of total length $Nt$ is $p = (\mathcal S + 1)/2$,
where
\begin{equation}
\label{eq_signal}
	{\mathcal S}
	= 1 - (\uvect n_0 \times \uvect n_1)^2 \sin^2\left(\tfrac{\omega_0 t}{2}\right) \sin^2  
		\left(\tfrac{\omega_1 t}{2}\right) 
		\frac{\sin^2 \left(N \phi/2\right)}{\cos^2 \left(\phi/2\right)}
\end{equation}
and
$\cos \phi = \cos\tfrac{\omega_0 t}{2} \cos \tfrac{\omega_1 t}{2}- \uvect n_0 \cdot \uvect n_1 \sin\tfrac{\omega_0 t}{2} \sin \tfrac{\omega_1 t}{2}$.
 This is the extension for arbitrary (even) $N$ pulses
of the well-known result for $N=1$ corresponding to Hahn spin echo
 \ccite{Childress2006}.
The factor of $N$ multiplying the small angle $\phi$
is responsible for the
enhanced signal that enables detection of weakly coupled
nuclear spins. 
We use \eq{eq_signal} with $N=8$ to fit all data below. %

{\it Experiment.}---We 
study NVs implanted in a diamond sample 
with naturally abundant spin-$\frac{1}{2}$ \C (1.1\% abundance). 
The NV electronic orbital ground state is a spin triplet which can be initialized in the $\ket{m_s = 0}$ state using green illumination, and read out via spin-dependent fluorescence \ccite{Childress2006}. 
The $\ket{m_s =\pm 1}$ degeneracy is lifted by a dc field applied using a permanent magnet. 
We coherently manipulate the NV spin using resonant microwaves delivered by a coplanar waveguide. The control pulse sequence used in this work is known as XY4-8 \ccite{deLange:2010ue}, 
and consists of $N$=8 $\pi$-pulses as shown in Fig.\,\ref{fig_1}b, phase alternated 
to mitigate microwave pulse errors. 

\begin{figure}[b]
\begin{center}
\includegraphics{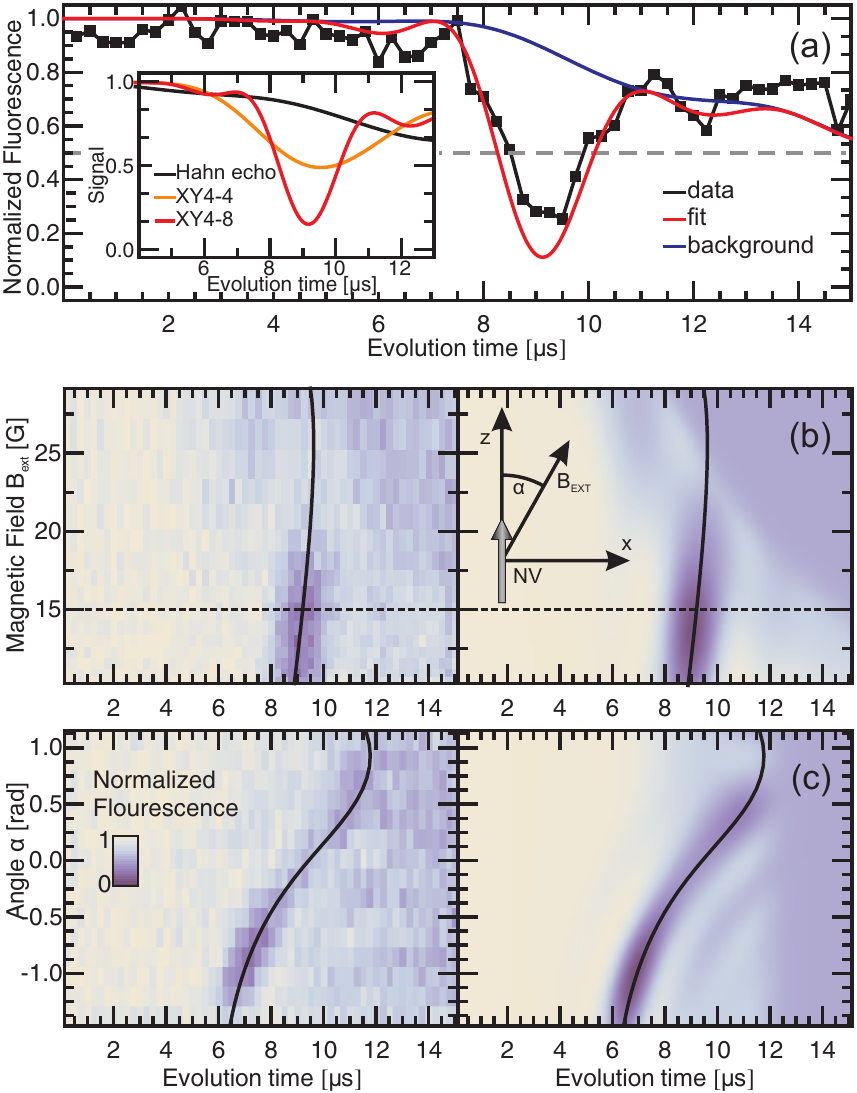}%
\end{center}
\caption{Coherent interaction of an NV-Center with a single \C detected using XY4-8. (a) The evolution time between $\pi$-pulses $t$ is swept with a fixed magnetic field of 15\,G, data (black) and fit (red). The blue line is the calculated response without the individual \C. The inset shows the calculated response to Hahn echo, XY4-4, and XY4-8, with the collapse resulting from the background spin bath kept fixed at the calculated response to XY4-8 to highlight the impact of the single \C.  (b) The fluorescence is plotted as a function of $t$ and magnetic field strength B, with magnetic field oriented along the NV $\hat{z}$ axis ($\alpha = 0$, as defined in the inset). The dashed line highlights the data shown in (a). $($c) Fluorescence as a function of $t$ and magnetic field orientation $\alpha$, with B = 19\,G. The black line displays the extracted dip position in both (b) and (c). Here the total coupling strength is $\left| \vect A \right|/2\pi \sim 125$\,kHz.} 
\label{fig_2}%
\end{figure}
%

%{\it Single {\rm \C} detection.}---
Detection of a single \C spin with a hyperfine coupling of $\left| \vect A \right|/2\pi \sim 125$\,kHz is shown in Fig.\,\ref{fig_2}a. 
The dip in the NV fluorescence signal at evolution time $t \approx 9$ $\mu$s results from the coherent interaction with a single \C spin and represents the signal. 
Importantly, this signal would go undetected using spin echo, as
shown in the inset of Fig.\,\ref{fig_2}a. In addition to the dip arising from the single \C, an overall decay is apparent and arises from the interaction with the bath of distant \Cs. The impact of this spin bath can be treated as an effective fluctuating ac magnetic field at the characteristic Larmor precession frequency \ccite{Childress2006}.  Consequently, the NV evolution will also periodically exhibit collapses and revivals in its coherence due to the bath \Cs, and the visible decay in Fig.\,\ref{fig_2}a is the onset of the first collapse (for the exact form of the collapse used in the fits, see \ccite{supp}).

To confirm that the observed feature arises from a single \C, we plot the measured signal as a function of applied magnetic field strength (Fig.\,\ref{fig_2}b) and orientation (Fig.\,\ref{fig_2}c). We measure the NV response in both possible sets of magnetic sublevels ($\ket 0 \leftrightarrow \ket 1$ and $\ket 0 \leftrightarrow \ket{-1}$, see \ccite{supp}) and the entire data set is simultaneously fitted to \eq{eq_signal}. The only free parameters are the hyperfine vector $\vect A$, defined relative to the NV axis and magnetic rotation axis (see \ccite{supp}), and two free parameters describing the strength and spectral width of the surrounding spin bath \ccite{supp}. 
The calculated signal using the extracted fit parameters is shown in the right panels, and is in good agreement with the data. The fit gives a total coupling strength of $\left| \vect A \right|/2\pi \sim125$\,kHz, which is less than the bare dephasing rate of this NV, $1/T_2^\ast = $ 400$\pm$16 kHz. $T_2^*$ is measured using a Ramsey sequence with no control $\pi$-pulses (measurement shown in \ccite{supp}), and arises from the bath spin configuration varying from measurement to measurement.
This demonstrates that we can simultaneously decouple the NV from the \C spin bath and still obtain a measurable signal from a target nuclear spin.

\begin{figure}[t]
\begin{center}
\includegraphics{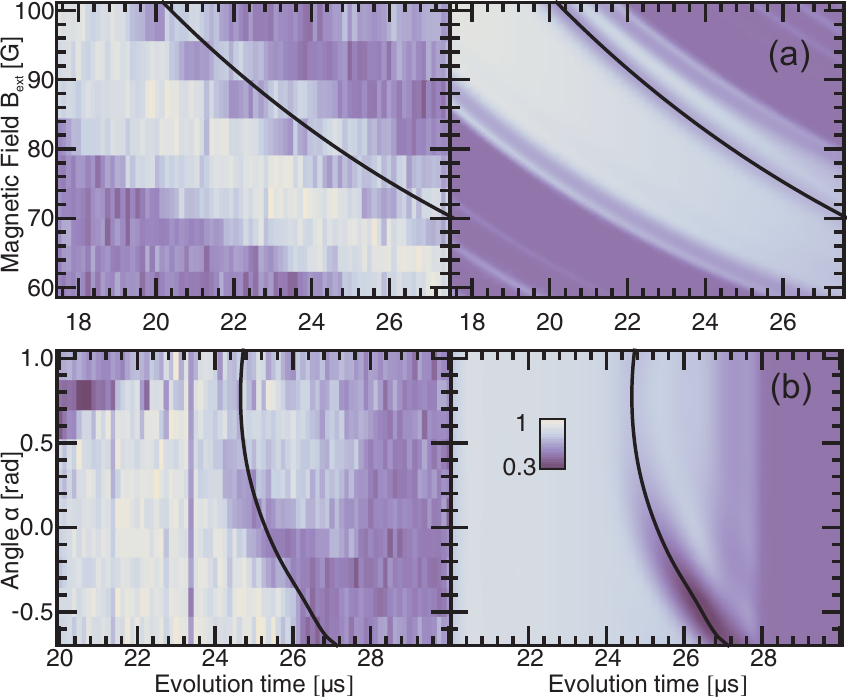}%
\end{center}
\caption{Detection of a distant \C with a XY4-8 sequence. As in Fig.\,\ref{fig_2} the NV fluorescence is shown as a function of (a) magnetic field strength B with $\alpha$ = 0.26\,rad, and $($b) magnetic field orientation $\alpha$, with B = 80\,G. The dip from the single \C is visible within a spin-bath revival centered at $2\cdot2\pi/\omega_0$.  The extracted coupling strength is $\left| \vect A \right|/2\pi \sim 47$\,kHz. } 
\label{fig_3}%
\end{figure}

%{\it Weakest observed coupling.}--- 
Figure \ref{fig_3} shows the measurement of the weakest coupled individual \C observed in this work.
This single spin produces no discernible signal for the experimental settings in Fig.\,\ref{fig_2} (measurement shown in \ccite{supp}), because
the weaker hyperfine coupling results in a dip at longer evolution times $t$, which occurs after the first collapse induced by the bath and is therefore suppressed. In order to measure this weakly coupled spin at very long evolution times we increase the magnetic field, moving a bath-induced revival into the range of interest. We thereby observe a characteristic dip in the bath revival resulting from an individual nuclear spin.
Again, this data is reproduced using our model calculations; the extracted coupling strength is $\left| \vect A \right|/2\pi \sim47$\,kHz, a factor of $\sim 4.6$ below the measured $1/T_2^\ast = 217\pm9$ kHz\ccite{supp}.

\begin{figure}[t]
\begin{center}
\includegraphics{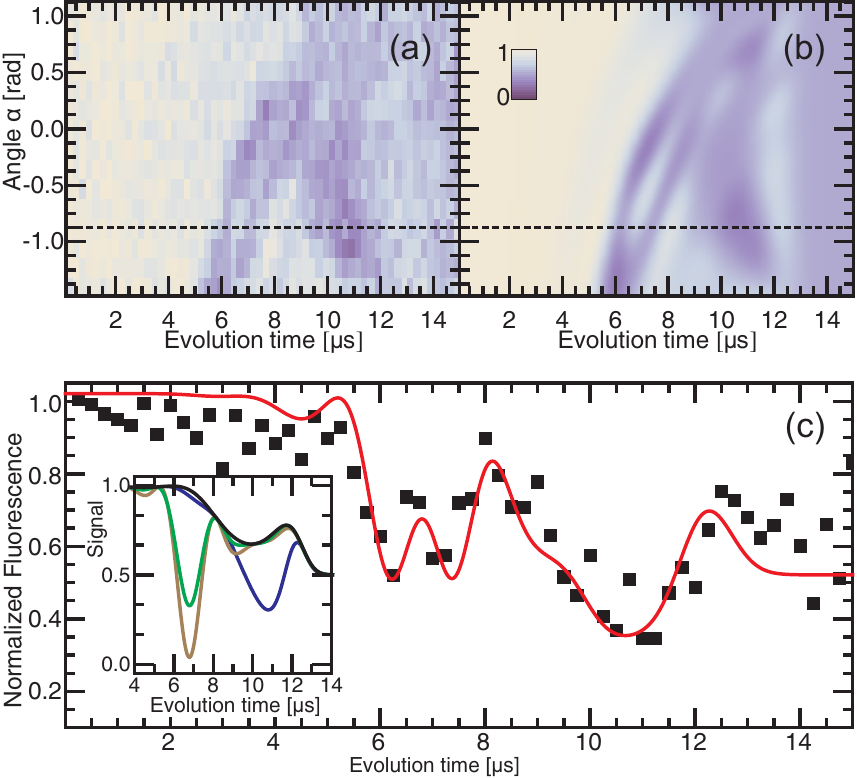}%
\end{center}
\caption{Simultaneous detection of three \C nuclei. (a) and (b) NV luminescence is shown vs. evolution time $t$ and magnetic field orientation $\alpha$ (data and fit), the applied field strength is 19\,G. The fit assumes three \C spins and yields unique individual coupling strengths, while fits assuming two or fewer \C spins could not reproduce the data. The NV response along the dashed line is shown in (c) black, red: data, fit. The traces in the inset display the calculated response of the NV interacting separately with each of the three coupled \Cs, and with the spin bath alone (black.) 
} 
\label{fig_4}%
\end{figure}

The coherent nature of the hyperfine interaction also allows us to observe the simultaneous impact of multiple \Cs coupled to a single NV. As the maximum mutual interaction strength of two \Cs on adjacent lattice sites is 2\,kHz\ccite{Zhao:2011fk}, it is negligible compared to the hyperfine interaction and to leading order we can treat the \Cs as independent. Consequently, the signal given by \eq{eq_signal} becomes a product of the individual contributions from each coupled spin, ${\mathcal S}=  \Pi_j {\mathcal S}_{j}$ \ccite{Childress2006}.
Figure\,\ref{fig_4} shows the impact of three \Cs on the evolution of an NV center. The double peak visible in the data results from two overlapping dips caused by two \C spins with similar coupling strengths. The peak in the middle reflects the product of two negative contributions and directly reflects the coherent nature of the interaction. This is emphasized in the inset of Fig.\,\ref{fig_4}c, where the calculated effect of all three individual \Cs on ${\mathcal S}$ is displayed separately on top of the bare signal resulting from the bath. The weakest coupled of the three \Cs has an extracted coupling strength of $\left| \vect A \right|/2\pi \sim64$\,kHz, a factor of $\sim 8$ below the measured $1/T_2^\ast= 560\pm60$ kHz \ccite{supp}.

{\it Discussion.}---We now turn to a discussion of our results and the limits of our technique. In contrast to the nearby \Cs detected in previous works\ccite{Childress2006, Smeltzer2011, Jacques2012}, the magnitudes of the hyperfine couplings observed here are consistent with a purely dipolar interaction. Assuming a point-dipole interaction, the measured couplings translate into NV-\C distances of $\sim$ 0.4-0.8 nm \ccite{supp}, the range expected for a probabilistic distribution of naturally abundant \C. However, ab initio calculations suggest that at this length scale the spatial extent of the electronic wavefunction can lead to significant contributions from the contact interaction\ccite{Gali2008, Gali2009, Smeltzer2011, Jacques2012}. Because our observed coupling strengths are an order of magnitude weaker than the contact interaction strengths predicted by these calculations\ccite{Gali2008, Gali2009}, we conclude that the dipolar interaction provides a minimum distance, and that the observed \Cs may be farther away than estimated. These considerations imply that our technique provides a tool to measure the spatial distribution of the electronic wave function beyond the limits of current theory, although a larger data set in the spirit of recent surveys of strongly coupled \Cs\ccite{Smeltzer2011, Jacques2012} would be required to provide the statistics necessary to identify individual distant \C lattice sites.

The sensitivity limits of our technique are determined by the NV spin decoherence. The hyperfine couplings $\left| \vect A \right|/2\pi$ extracted in this work exceed the decoherence rate $1/T_2$ obtained in a Hahn spin echo sequence, and in principle the \Cs could therefore be detected using only a single $\pi$-pulse\ccite{supp}. However, this would require precise optimization of the magnetic field strength and orientation, which is not practical in a typical experiment. One advantage of extended pulse sequences is to relax these conditions and greatly simplify the identification of nuclear spins whose hyperfine coupling with the NV spin are initially unknown. As a result, using the presented technique we were able to detect and characterize the closest \C for every NV investigated. 

The use of dynamical decoupling sequences with multiple $\pi$-pulses has the added benefit of decoupling the NV spin from the surrounding environment, increasing its coherence time from $T_2$ to $T_2^{\rm eff} = T_2 N^{2/3}$ and thereby potentially improving the sensitivity beyond the limit set by $T_2$. Increasing the number of $\pi$-pulses for the same total evolution time requires higher magnetic fields to ensure that the sequence remains synchronized with the evolution of the target nuclear spin. We note that in the limit $\left| \mu_n\vect B \right| \gg \left| \vect A \right|$ a weakly coupled single \C has the same Larmor precession frequency as the surrounding nuclear spin bath and cannot be isolated. Therefore $\left| \vect A \right|/2\pi > 1/T_2$ is the limit for the detection of individual \C nuclear spins in a sample of natural isotopic abundance.  
However, for applications involving the detection of 
spins with a different gyromagnetic ratio than the \C bath we find that the
sensitivity limits of our technique improve with number of $\pi$-pulses as $N^{4/3}$\ccite{supp}. 
As a result, our approach could potentially be used to detect individual spins inside or outside the diamond lattice up to the ultimate limit $\left| \vect A \right|/2\pi > 1/2T_1$, where $1/T_1$ is the NV spin relaxation rate. At room temperature $1/2T_1$ can be as low as $\sim$100 Hz \ccite{Jarmola:2011wf, Bala:2009ub}, corresponding to the dipolar coupling between an NV and a proton spin at a distance of $\sim$7-9 nm, which is of great interest for single spin MRI applications\ccite{Rugar:2004wi, Degen:2009wa,Hammel2007}. 

In conclusion, we used coherently controlled single NV electronic spins to detect distant nuclear spins. Extended pulse sequences allow for detection of individual weakly coupled \C spins that would otherwise be unresolvable.  We also showed that the simultaneous detection of several distant \Cs is possible, even within an environment consisting of a large number of spins with the same gyromagnetic ratio. Our technique allows for sensitive, coherent measurements of the nuclear spin environment of a single electronic spin; moreover, extensions of this approach could be used to exploit the nuclear spin environment as a resource to extend the size of controllable multi-spin quantum registers. Potential novel applications range from information storage to environment-assisted sensing and single spin MRI. 

{\it Authors' Note.}---Following submission of this work two complementary studies appeared\ccite{Taminiau2012,Zhao2012} in which the technique demonstrated in this work was used to resolve individual \C nuclei weakly coupled to a single NV in the high magnetic field regime\ccite{Taminiau2012}, and in isotopically purified diamond with depleted \C nuclei \ccite{Zhao2012}. 

{\it Acknowledgements.}---We thank Jack Harris, Lillian Childress, Adam Gali, Ronald Hanson, and Tim Taminiau for stimulating discussions, and Alexander Zibrov for early contributions to the experiment. This work was supported in part by the National Science Foundation (NSF), the Center for Ultracold Atoms (CUA), AFSOR MURI, DARPA QUASAR, EU DIAMANT, and the Packard Foundation. S.K. acknowledges support by the DoD through the NDSEG Program, and the NSF through the NSFGRP under Grant No. DGE-1144152. Q.U. acknowledges support from Deutschen Forschungsgemeinschaft. S.D.B. acknowledges support from NSERC of Canada and ITAMP.

\normalsize{$^\dagger$These authors contributed equally.} \\

\normalsize{$^\ast$To whom correspondence should be addressed; E-mail:  quirin@physics.harvard.edu}    

\begin{thebibliography}{30}
\expandafter\ifx\csname natexlab\endcsname\relax\def\natexlab#1{#1}\fi
\expandafter\ifx\csname bibnamefont\endcsname\relax
  \def\bibnamefont#1{#1}\fi
\expandafter\ifx\csname bibfnamefont\endcsname\relax
  \def\bibfnamefont#1{#1}\fi
\expandafter\ifx\csname citenamefont\endcsname\relax
  \def\citenamefont#1{#1}\fi
\expandafter\ifx\csname url\endcsname\relax
  \def\url#1{\texttt{#1}}\fi
\expandafter\ifx\csname urlprefix\endcsname\relax\def\urlprefix{URL }\fi
\providecommand{\bibinfo}[2]{#2}
\providecommand{\eprint}[2][]{\url{#2}}

\bibitem[{\citenamefont{Rugar et~al.}(2004)\citenamefont{Rugar, Budakian,
  Mamin, and Chui}}]{Rugar:2004wi}
\bibinfo{author}{\bibfnamefont{D.}~\bibnamefont{Rugar}},
  \bibinfo{author}{\bibfnamefont{R.}~\bibnamefont{Budakian}},
  \bibinfo{author}{\bibfnamefont{H.}~\bibnamefont{Mamin}}, \bibnamefont{and}
  \bibinfo{author}{\bibfnamefont{B.}~\bibnamefont{Chui}},
  \bibinfo{journal}{Nature} \textbf{\bibinfo{volume}{430}},
  \bibinfo{pages}{329} (\bibinfo{year}{2004}).

\bibitem[{\citenamefont{Degen et~al.}(2009)\citenamefont{Degen, Poggio, Mamin,
  Rettner, and Rugar}}]{Degen:2009wa}
\bibinfo{author}{\bibfnamefont{C.~L.} \bibnamefont{Degen}},
  \bibinfo{author}{\bibfnamefont{M.}~\bibnamefont{Poggio}},
  \bibinfo{author}{\bibfnamefont{H.~J.} \bibnamefont{Mamin}},
  \bibinfo{author}{\bibfnamefont{C.~T.} \bibnamefont{Rettner}},
  \bibnamefont{and} \bibinfo{author}{\bibfnamefont{D.}~\bibnamefont{Rugar}},
  \bibinfo{journal}{Proceedings of the National Academy of Sciences}
  \textbf{\bibinfo{volume}{106}}, \bibinfo{pages}{1313} (\bibinfo{year}{2009}).

\bibitem[{\citenamefont{Hammel and Pelekhov}(2007)}]{Hammel2007}
\bibinfo{author}{\bibfnamefont{P.~C.} \bibnamefont{Hammel}} \bibnamefont{and}
  \bibinfo{author}{\bibfnamefont{D.~V.} \bibnamefont{Pelekhov}},
  \emph{\bibinfo{title}{Handbook of Magnetism and Advanced Magnetic Materials}}
  (\bibinfo{publisher}{Wiley}, \bibinfo{address}{New York},
  \bibinfo{year}{2007}), vol.~\bibinfo{volume}{5}, chap.~\bibinfo{chapter}{4}.

\bibitem[{\citenamefont{Jelezko et~al.}(2004)\citenamefont{Jelezko, Gaebel,
  Popa, Domhan, Gruber, and Wrachtrup}}]{Jelezko2004}
\bibinfo{author}{\bibfnamefont{F.}~\bibnamefont{Jelezko}},
  \bibinfo{author}{\bibfnamefont{T.}~\bibnamefont{Gaebel}},
  \bibinfo{author}{\bibfnamefont{I.}~\bibnamefont{Popa}},
  \bibinfo{author}{\bibfnamefont{M.}~\bibnamefont{Domhan}},
  \bibinfo{author}{\bibfnamefont{A.}~\bibnamefont{Gruber}}, \bibnamefont{and}
  \bibinfo{author}{\bibfnamefont{J.}~\bibnamefont{Wrachtrup}},
  \bibinfo{journal}{Physical Review Letters} \textbf{\bibinfo{volume}{93}},
  \bibinfo{pages}{130501} (\bibinfo{year}{2004}).

\bibitem[{\citenamefont{Childress et~al.}(2006)\citenamefont{Childress,
  {Gurudev Dutt}, Taylor, Zibrov, Jelezko, Wrachtrup, Hemmer, and
  Lukin}}]{Childress2006}
\bibinfo{author}{\bibfnamefont{L.}~\bibnamefont{Childress}},
  \bibinfo{author}{\bibfnamefont{M.~V.} \bibnamefont{{Gurudev Dutt}}},
  \bibinfo{author}{\bibfnamefont{J.~M.} \bibnamefont{Taylor}},
  \bibinfo{author}{\bibfnamefont{A.~S.} \bibnamefont{Zibrov}},
  \bibinfo{author}{\bibfnamefont{F.}~\bibnamefont{Jelezko}},
  \bibinfo{author}{\bibfnamefont{J.}~\bibnamefont{Wrachtrup}},
  \bibinfo{author}{\bibfnamefont{P.~R.} \bibnamefont{Hemmer}},
  \bibnamefont{and} \bibinfo{author}{\bibfnamefont{M.~D.} \bibnamefont{Lukin}},
  \bibinfo{journal}{Science} \textbf{\bibinfo{volume}{314}},
  \bibinfo{pages}{281} (\bibinfo{year}{2006}).

\bibitem[{\citenamefont{Hanson et~al.}(2008)\citenamefont{Hanson, Dobrovitski,
  Feiguin, Gywat, and Awschalom}}]{Hanson2008}
\bibinfo{author}{\bibfnamefont{R.}~\bibnamefont{Hanson}},
  \bibinfo{author}{\bibfnamefont{V.~V.} \bibnamefont{Dobrovitski}},
  \bibinfo{author}{\bibfnamefont{A.~E.} \bibnamefont{Feiguin}},
  \bibinfo{author}{\bibfnamefont{O.}~\bibnamefont{Gywat}}, \bibnamefont{and}
  \bibinfo{author}{\bibfnamefont{D.~D.} \bibnamefont{Awschalom}},
  \bibinfo{journal}{Science} \textbf{\bibinfo{volume}{320}},
  \bibinfo{pages}{352} (\bibinfo{year}{2008}).

\bibitem[{\citenamefont{{Gurudev Dutt} et~al.}(2007)\citenamefont{{Gurudev
  Dutt}, Childress, Jiang, Togan, Maze, Jelezko, Zibrov, Hemmer, and
  Lukin}}]{Dutt2007}
\bibinfo{author}{\bibfnamefont{M.~V.} \bibnamefont{{Gurudev Dutt}}},
  \bibinfo{author}{\bibfnamefont{L.}~\bibnamefont{Childress}},
  \bibinfo{author}{\bibfnamefont{L.}~\bibnamefont{Jiang}},
  \bibinfo{author}{\bibfnamefont{E.}~\bibnamefont{Togan}},
  \bibinfo{author}{\bibfnamefont{J.}~\bibnamefont{Maze}},
  \bibinfo{author}{\bibfnamefont{F.}~\bibnamefont{Jelezko}},
  \bibinfo{author}{\bibfnamefont{A.~S.} \bibnamefont{Zibrov}},
  \bibinfo{author}{\bibfnamefont{P.~R.} \bibnamefont{Hemmer}},
  \bibnamefont{and} \bibinfo{author}{\bibfnamefont{M.~D.} \bibnamefont{Lukin}},
  \bibinfo{journal}{Science} \textbf{\bibinfo{volume}{316}},
  \bibinfo{pages}{1312} (\bibinfo{year}{2007}).

\bibitem[{\citenamefont{Jiang et~al.}(2009)\citenamefont{Jiang, Hodges, Maze,
  Maurer, Taylor, Cory, Hemmer, Walsworth, Yacoby, Zibrov et~al.}}]{Jiang2009}
\bibinfo{author}{\bibfnamefont{L.}~\bibnamefont{Jiang}},
  \bibinfo{author}{\bibfnamefont{J.~S.} \bibnamefont{Hodges}},
  \bibinfo{author}{\bibfnamefont{J.~R.} \bibnamefont{Maze}},
  \bibinfo{author}{\bibfnamefont{P.}~\bibnamefont{Maurer}},
  \bibinfo{author}{\bibfnamefont{J.~M.} \bibnamefont{Taylor}},
  \bibinfo{author}{\bibfnamefont{D.~G.} \bibnamefont{Cory}},
  \bibinfo{author}{\bibfnamefont{P.~R.} \bibnamefont{Hemmer}},
  \bibinfo{author}{\bibfnamefont{R.~L.} \bibnamefont{Walsworth}},
  \bibinfo{author}{\bibfnamefont{A.}~\bibnamefont{Yacoby}},
  \bibinfo{author}{\bibfnamefont{A.~S.} \bibnamefont{Zibrov}},
  \bibnamefont{et~al.}, \bibinfo{journal}{Science}
  \textbf{\bibinfo{volume}{326}}, \bibinfo{pages}{267} (\bibinfo{year}{2009}).

\bibitem[{\citenamefont{van~der Sar et~al.}(2012)\citenamefont{van~der Sar,
  Wang, Blok, Bernien, Taminiau, Toyli, Lidar, Awschalom, Hanson, and
  Dobrovitski}}]{Sar:2012fk}
\bibinfo{author}{\bibfnamefont{T.}~\bibnamefont{van~der Sar}},
  \bibinfo{author}{\bibfnamefont{Z.~H.} \bibnamefont{Wang}},
  \bibinfo{author}{\bibfnamefont{M.~S.} \bibnamefont{Blok}},
  \bibinfo{author}{\bibfnamefont{H.}~\bibnamefont{Bernien}},
  \bibinfo{author}{\bibfnamefont{T.~H.} \bibnamefont{Taminiau}},
  \bibinfo{author}{\bibfnamefont{D.~M.} \bibnamefont{Toyli}},
  \bibinfo{author}{\bibfnamefont{D.~A.} \bibnamefont{Lidar}},
  \bibinfo{author}{\bibfnamefont{D.~D.} \bibnamefont{Awschalom}},
  \bibinfo{author}{\bibfnamefont{R.}~\bibnamefont{Hanson}}, \bibnamefont{and}
  \bibinfo{author}{\bibfnamefont{V.~V.} \bibnamefont{Dobrovitski}},
  \bibinfo{journal}{Nature} \textbf{\bibinfo{volume}{484}}, \bibinfo{pages}{82}
  (\bibinfo{year}{2012}).

\bibitem[{\citenamefont{Neumann et~al.}(2010)\citenamefont{Neumann, Beck,
  Steiner, Rempp, Fedder, Hemmer, Wrachtrup, and Jelezko}}]{Neumann2010}
\bibinfo{author}{\bibfnamefont{P.}~\bibnamefont{Neumann}},
  \bibinfo{author}{\bibfnamefont{J.}~\bibnamefont{Beck}},
  \bibinfo{author}{\bibfnamefont{M.}~\bibnamefont{Steiner}},
  \bibinfo{author}{\bibfnamefont{F.}~\bibnamefont{Rempp}},
  \bibinfo{author}{\bibfnamefont{H.}~\bibnamefont{Fedder}},
  \bibinfo{author}{\bibfnamefont{P.~R.} \bibnamefont{Hemmer}},
  \bibinfo{author}{\bibfnamefont{J.}~\bibnamefont{Wrachtrup}},
  \bibnamefont{and} \bibinfo{author}{\bibfnamefont{F.}~\bibnamefont{Jelezko}},
  \bibinfo{journal}{Science} \textbf{\bibinfo{volume}{329}},
  \bibinfo{pages}{542} (\bibinfo{year}{2010}).

\bibitem[{\citenamefont{Robledo et~al.}(2011)\citenamefont{Robledo, Childress,
  Bernien, Hensen, Alkemade, and Hanson}}]{Childress:2011wv}
\bibinfo{author}{\bibfnamefont{L.}~\bibnamefont{Robledo}},
  \bibinfo{author}{\bibfnamefont{L.}~\bibnamefont{Childress}},
  \bibinfo{author}{\bibfnamefont{H.}~\bibnamefont{Bernien}},
  \bibinfo{author}{\bibfnamefont{B.}~\bibnamefont{Hensen}},
  \bibinfo{author}{\bibfnamefont{P.~F.~A.} \bibnamefont{Alkemade}},
  \bibnamefont{and} \bibinfo{author}{\bibfnamefont{R.}~\bibnamefont{Hanson}},
  \bibinfo{journal}{Nature} \textbf{\bibinfo{volume}{477}},
  \bibinfo{pages}{574} (\bibinfo{year}{2011}).

\bibitem[{\citenamefont{Maurer et~al.}(2012)\citenamefont{Maurer, Kucsko,
  Latta, Jiang, Yao, Bennett, Pastawski, Hunger, Chisholm, Markham
  et~al.}}]{Maurer2012}
\bibinfo{author}{\bibfnamefont{P.~C.} \bibnamefont{Maurer}},
  \bibinfo{author}{\bibfnamefont{G.}~\bibnamefont{Kucsko}},
  \bibinfo{author}{\bibfnamefont{C.}~\bibnamefont{Latta}},
  \bibinfo{author}{\bibfnamefont{L.}~\bibnamefont{Jiang}},
  \bibinfo{author}{\bibfnamefont{N.~Y.} \bibnamefont{Yao}},
  \bibinfo{author}{\bibfnamefont{S.~D.} \bibnamefont{Bennett}},
  \bibinfo{author}{\bibfnamefont{F.}~\bibnamefont{Pastawski}},
  \bibinfo{author}{\bibfnamefont{D.}~\bibnamefont{Hunger}},
  \bibinfo{author}{\bibfnamefont{N.}~\bibnamefont{Chisholm}},
  \bibinfo{author}{\bibfnamefont{M.}~\bibnamefont{Markham}},
  \bibnamefont{et~al.}, \bibinfo{journal}{In Press (Science)}
  (\bibinfo{year}{2012}).

\bibitem[{\citenamefont{Taylor et~al.}(2008)\citenamefont{Taylor, Cappellaro,
  Childress, Jiang, Budker, Hemmer, Yacoby, Walsworth, and
  Lukin}}]{Taylor:2008ub}
\bibinfo{author}{\bibfnamefont{J.~M.} \bibnamefont{Taylor}},
  \bibinfo{author}{\bibfnamefont{P.}~\bibnamefont{Cappellaro}},
  \bibinfo{author}{\bibfnamefont{L.}~\bibnamefont{Childress}},
  \bibinfo{author}{\bibfnamefont{L.}~\bibnamefont{Jiang}},
  \bibinfo{author}{\bibfnamefont{D.}~\bibnamefont{Budker}},
  \bibinfo{author}{\bibfnamefont{P.~R.} \bibnamefont{Hemmer}},
  \bibinfo{author}{\bibfnamefont{A.}~\bibnamefont{Yacoby}},
  \bibinfo{author}{\bibfnamefont{R.}~\bibnamefont{Walsworth}},
  \bibnamefont{and} \bibinfo{author}{\bibfnamefont{M.~D.} \bibnamefont{Lukin}},
  \bibinfo{journal}{Nature Physics} \textbf{\bibinfo{volume}{4}},
  \bibinfo{pages}{810} (\bibinfo{year}{2008}).

\bibitem[{\citenamefont{de~Lange et~al.}(2010)\citenamefont{de~Lange, Wang,
  Rist{\`e}, Dobrovitski, and Hanson}}]{deLange:2010ue}
\bibinfo{author}{\bibfnamefont{G.}~\bibnamefont{de~Lange}},
  \bibinfo{author}{\bibfnamefont{Z.~H.} \bibnamefont{Wang}},
  \bibinfo{author}{\bibfnamefont{D.}~\bibnamefont{Rist{\`e}}},
  \bibinfo{author}{\bibfnamefont{V.~V.} \bibnamefont{Dobrovitski}},
  \bibnamefont{and} \bibinfo{author}{\bibfnamefont{R.}~\bibnamefont{Hanson}},
  \bibinfo{journal}{Science} \textbf{\bibinfo{volume}{330}},
  \bibinfo{pages}{60} (\bibinfo{year}{2010}).

\bibitem[{\citenamefont{Hall et~al.}(2010)\citenamefont{Hall, Hill, Cole, and
  Hollenberg}}]{Hall:2010vs}
\bibinfo{author}{\bibfnamefont{L.~T.} \bibnamefont{Hall}},
  \bibinfo{author}{\bibfnamefont{C.~D.} \bibnamefont{Hill}},
  \bibinfo{author}{\bibfnamefont{J.~H.} \bibnamefont{Cole}}, \bibnamefont{and}
  \bibinfo{author}{\bibfnamefont{L.~C.~L.} \bibnamefont{Hollenberg}},
  \bibinfo{journal}{Physical Review B} \textbf{\bibinfo{volume}{82}},
  \bibinfo{pages}{045208} (\bibinfo{year}{2010}).

\bibitem[{\citenamefont{Naydenov et~al.}(2011)\citenamefont{Naydenov, Dolde,
  Hall, Shin, Fedder, Hollenberg, Jelezko, and Wrachtrup}}]{Naydenov:2011uz}
\bibinfo{author}{\bibfnamefont{B.}~\bibnamefont{Naydenov}},
  \bibinfo{author}{\bibfnamefont{F.}~\bibnamefont{Dolde}},
  \bibinfo{author}{\bibfnamefont{L.~T.} \bibnamefont{Hall}},
  \bibinfo{author}{\bibfnamefont{C.}~\bibnamefont{Shin}},
  \bibinfo{author}{\bibfnamefont{H.}~\bibnamefont{Fedder}},
  \bibinfo{author}{\bibfnamefont{L.~C.~L.} \bibnamefont{Hollenberg}},
  \bibinfo{author}{\bibfnamefont{F.}~\bibnamefont{Jelezko}}, \bibnamefont{and}
  \bibinfo{author}{\bibfnamefont{J.}~\bibnamefont{Wrachtrup}},
  \bibinfo{journal}{Physical Review B} \textbf{\bibinfo{volume}{83}},
  \bibinfo{pages}{081201} (\bibinfo{year}{2011}).

\bibitem[{\citenamefont{de~Lange et~al.}(2011)\citenamefont{de~Lange, Riste,
  Dobrovitski, and Hanson}}]{deLange:2011wm}
\bibinfo{author}{\bibfnamefont{G.}~\bibnamefont{de~Lange}},
  \bibinfo{author}{\bibfnamefont{D.}~\bibnamefont{Riste}},
  \bibinfo{author}{\bibfnamefont{V.~V.} \bibnamefont{Dobrovitski}},
  \bibnamefont{and} \bibinfo{author}{\bibfnamefont{R.}~\bibnamefont{Hanson}},
  \bibinfo{journal}{Physical Review Letters} \textbf{\bibinfo{volume}{106}},
  \bibinfo{pages}{80802} (\bibinfo{year}{2011}).

\bibitem[{\citenamefont{Ryan et~al.}(2010)\citenamefont{Ryan, Hodges, and
  Cory}}]{Ryan2010}
\bibinfo{author}{\bibfnamefont{C.~A.} \bibnamefont{Ryan}},
  \bibinfo{author}{\bibfnamefont{J.~S.} \bibnamefont{Hodges}},
  \bibnamefont{and} \bibinfo{author}{\bibfnamefont{D.~G.} \bibnamefont{Cory}},
  \bibinfo{journal}{Physical Review Letters} \textbf{\bibinfo{volume}{105}},
  \bibinfo{pages}{1} (\bibinfo{year}{2010}), ISSN \bibinfo{issn}{0031-9007}.

\bibitem[{\citenamefont{{Wang} et~al.}(2012)\citenamefont{{Wang}, {de Lange},
  {Rist{\`e}}, {Hanson}, and {Dobrovitski}}}]{Dobrovitski2012}
\bibinfo{author}{\bibfnamefont{Z.-H.} \bibnamefont{{Wang}}},
  \bibinfo{author}{\bibfnamefont{G.}~\bibnamefont{{de Lange}}},
  \bibinfo{author}{\bibfnamefont{D.}~\bibnamefont{{Rist{\`e}}}},
  \bibinfo{author}{\bibfnamefont{R.}~\bibnamefont{{Hanson}}}, \bibnamefont{and}
  \bibinfo{author}{\bibfnamefont{V.~V.} \bibnamefont{{Dobrovitski}}},
  \bibinfo{journal}{\prb} \textbf{\bibinfo{volume}{85}}, \bibinfo{eid}{155204}
  (\bibinfo{year}{2012}), \eprint{1202.0462}.

\bibitem[{\citenamefont{Gali et~al.}(2008)\citenamefont{Gali, Fyta, and
  Kaxiras}}]{Gali2008}
\bibinfo{author}{\bibfnamefont{A.}~\bibnamefont{Gali}},
  \bibinfo{author}{\bibfnamefont{M.}~\bibnamefont{Fyta}}, \bibnamefont{and}
  \bibinfo{author}{\bibfnamefont{E.}~\bibnamefont{Kaxiras}},
  \bibinfo{journal}{Physical Review B} \textbf{\bibinfo{volume}{77}},
  \bibinfo{pages}{1} (\bibinfo{year}{2008}).

\bibitem[{\citenamefont{Gali}(2009)}]{Gali2009}
\bibinfo{author}{\bibfnamefont{A.}~\bibnamefont{Gali}}, \bibinfo{journal}{Phys.
  Rev. B} \textbf{\bibinfo{volume}{80}}, \bibinfo{pages}{241204}
  (\bibinfo{year}{2009}).

\bibitem[{\citenamefont{Smeltzer et~al.}(2011)\citenamefont{Smeltzer,
  Childress, and Gali}}]{Smeltzer2011}
\bibinfo{author}{\bibfnamefont{B.}~\bibnamefont{Smeltzer}},
  \bibinfo{author}{\bibfnamefont{L.}~\bibnamefont{Childress}},
  \bibnamefont{and} \bibinfo{author}{\bibfnamefont{A.}~\bibnamefont{Gali}},
  \bibinfo{journal}{New Journal of Physics} \textbf{\bibinfo{volume}{13}},
  \bibinfo{pages}{025021} (\bibinfo{year}{2011}).

\bibitem[{\citenamefont{Dr\'eau et~al.}(2012)\citenamefont{Dr\'eau, Maze,
  Lesik, Roch, and Jacques}}]{Jacques2012}
\bibinfo{author}{\bibfnamefont{A.}~\bibnamefont{Dr\'eau}},
  \bibinfo{author}{\bibfnamefont{J.-R.} \bibnamefont{Maze}},
  \bibinfo{author}{\bibfnamefont{M.}~\bibnamefont{Lesik}},
  \bibinfo{author}{\bibfnamefont{J.-F.} \bibnamefont{Roch}}, \bibnamefont{and}
  \bibinfo{author}{\bibfnamefont{V.}~\bibnamefont{Jacques}},
  \bibinfo{journal}{Phys. Rev. B} \textbf{\bibinfo{volume}{85}},
  \bibinfo{pages}{134107} (\bibinfo{year}{2012}).

\bibitem[{\citenamefont{Maze et~al.}(2008)\citenamefont{Maze, Stanwix, Hodges,
  Hong, Taylor, Cappellaro, Jiang, {Gurudev Dutt}, Togan, Zibrov
  et~al.}}]{Maze:2008ws}
\bibinfo{author}{\bibfnamefont{J.~R.} \bibnamefont{Maze}},
  \bibinfo{author}{\bibfnamefont{P.~L.} \bibnamefont{Stanwix}},
  \bibinfo{author}{\bibfnamefont{J.~S.} \bibnamefont{Hodges}},
  \bibinfo{author}{\bibfnamefont{S.}~\bibnamefont{Hong}},
  \bibinfo{author}{\bibfnamefont{J.~M.} \bibnamefont{Taylor}},
  \bibinfo{author}{\bibfnamefont{P.}~\bibnamefont{Cappellaro}},
  \bibinfo{author}{\bibfnamefont{L.}~\bibnamefont{Jiang}},
  \bibinfo{author}{\bibfnamefont{M.~V.} \bibnamefont{{Gurudev Dutt}}},
  \bibinfo{author}{\bibfnamefont{E.}~\bibnamefont{Togan}},
  \bibinfo{author}{\bibfnamefont{A.~S.} \bibnamefont{Zibrov}},
  \bibnamefont{et~al.}, \bibinfo{journal}{Nature}
  \textbf{\bibinfo{volume}{455}}, \bibinfo{pages}{644} (\bibinfo{year}{2008}).

\bibitem[{\citenamefont{{Supplemental material}}(2012)}]{supp}
\bibinfo{author}{\bibnamefont{{Supplemental material}}} (\bibinfo{year}{2012}).

\bibitem[{\citenamefont{Zhao et~al.}(2011)\citenamefont{Zhao, Hu, Ho, Wan, and
  Liu}}]{Zhao:2011fk}
\bibinfo{author}{\bibfnamefont{N.}~\bibnamefont{Zhao}},
  \bibinfo{author}{\bibfnamefont{J.-L.} \bibnamefont{Hu}},
  \bibinfo{author}{\bibfnamefont{S.-W.} \bibnamefont{Ho}},
  \bibinfo{author}{\bibfnamefont{J.~T.~K.} \bibnamefont{Wan}},
  \bibnamefont{and} \bibinfo{author}{\bibfnamefont{R.~B.} \bibnamefont{Liu}},
  \bibinfo{journal}{Nature Nanotechnology} \textbf{\bibinfo{volume}{6}},
  \bibinfo{pages}{242} (\bibinfo{year}{2011}).

\bibitem[{\citenamefont{Jarmola et~al.}(2011)\citenamefont{Jarmola, Acosta,
  Jensen, Chemerisov, and Budker}}]{Jarmola:2011wf}
\bibinfo{author}{\bibfnamefont{A.}~\bibnamefont{Jarmola}},
  \bibinfo{author}{\bibfnamefont{V.~M.} \bibnamefont{Acosta}},
  \bibinfo{author}{\bibfnamefont{K.}~\bibnamefont{Jensen}},
  \bibinfo{author}{\bibfnamefont{S.}~\bibnamefont{Chemerisov}},
  \bibnamefont{and} \bibinfo{author}{\bibfnamefont{D.}~\bibnamefont{Budker}},
  \bibinfo{journal}{arXiv.org} \textbf{\bibinfo{volume}{cond-mat.mtrl-sci}}
  (\bibinfo{year}{2011}).

\bibitem[{\citenamefont{Balasubramanian
  et~al.}(2009)\citenamefont{Balasubramanian, Neumann, Twitchen, Markham,
  Kolesov, Mizuochi, Isoya, Achard, Beck, Tissler et~al.}}]{Bala:2009ub}
\bibinfo{author}{\bibfnamefont{G.}~\bibnamefont{Balasubramanian}},
  \bibinfo{author}{\bibfnamefont{P.}~\bibnamefont{Neumann}},
  \bibinfo{author}{\bibfnamefont{D.}~\bibnamefont{Twitchen}},
  \bibinfo{author}{\bibfnamefont{M.}~\bibnamefont{Markham}},
  \bibinfo{author}{\bibfnamefont{R.}~\bibnamefont{Kolesov}},
  \bibinfo{author}{\bibfnamefont{N.}~\bibnamefont{Mizuochi}},
  \bibinfo{author}{\bibfnamefont{J.}~\bibnamefont{Isoya}},
  \bibinfo{author}{\bibfnamefont{J.}~\bibnamefont{Achard}},
  \bibinfo{author}{\bibfnamefont{J.}~\bibnamefont{Beck}},
  \bibinfo{author}{\bibfnamefont{J.}~\bibnamefont{Tissler}},
  \bibnamefont{et~al.}, \bibinfo{journal}{Nature Materials}
  \textbf{\bibinfo{volume}{8}}, \bibinfo{pages}{383} (\bibinfo{year}{2009}).

\bibitem[{\citenamefont{{Taminiau} et~al.}(2012)\citenamefont{{Taminiau},
  {Wagenaar}, {van der Sar}, {Jelezko}, {Dobrovitski}, and
  {Hanson}}}]{Taminiau2012}
\bibinfo{author}{\bibfnamefont{T.~H.} \bibnamefont{{Taminiau}}},
  \bibinfo{author}{\bibfnamefont{J.~J.~T.} \bibnamefont{{Wagenaar}}},
  \bibinfo{author}{\bibfnamefont{T.}~\bibnamefont{{van der Sar}}},
  \bibinfo{author}{\bibfnamefont{F.}~\bibnamefont{{Jelezko}}},
  \bibinfo{author}{\bibfnamefont{V.~V.} \bibnamefont{{Dobrovitski}}},
  \bibnamefont{and} \bibinfo{author}{\bibfnamefont{R.}~\bibnamefont{{Hanson}}},
  \bibinfo{journal}{ArXiv e-prints}  (\bibinfo{year}{2012}),
  \eprint{1205.4128}.

\bibitem[{\citenamefont{{Zhao} et~al.}(2012)\citenamefont{{Zhao}, {Honert},
  {Schmid}, {Isoya}, {Markham}, {Twitchen}, {Jelezko}, {Liu}, {Fedder}, and
  {Wrachtrup}}}]{Zhao2012}
\bibinfo{author}{\bibfnamefont{N.}~\bibnamefont{{Zhao}}},
  \bibinfo{author}{\bibfnamefont{J.}~\bibnamefont{{Honert}}},
  \bibinfo{author}{\bibfnamefont{B.}~\bibnamefont{{Schmid}}},
  \bibinfo{author}{\bibfnamefont{J.}~\bibnamefont{{Isoya}}},
  \bibinfo{author}{\bibfnamefont{M.}~\bibnamefont{{Markham}}},
  \bibinfo{author}{\bibfnamefont{D.}~\bibnamefont{{Twitchen}}},
  \bibinfo{author}{\bibfnamefont{F.}~\bibnamefont{{Jelezko}}},
  \bibinfo{author}{\bibfnamefont{R.-B.} \bibnamefont{{Liu}}},
  \bibinfo{author}{\bibfnamefont{H.}~\bibnamefont{{Fedder}}}, \bibnamefont{and}
  \bibinfo{author}{\bibfnamefont{J.}~\bibnamefont{{Wrachtrup}}},
  \bibinfo{journal}{ArXiv e-prints}  (\bibinfo{year}{2012}),
  \eprint{1204.6513}.

\end{thebibliography}
\end{document}

\title{Supplemental material for ``Sensing distant nuclear spins with an single electron spin''}

\maketitle   
 
\tableofcontents

\section{System Hamiltonian and Signal Derivation}

\begin{figure}[b]
\begin{center}
\includegraphics{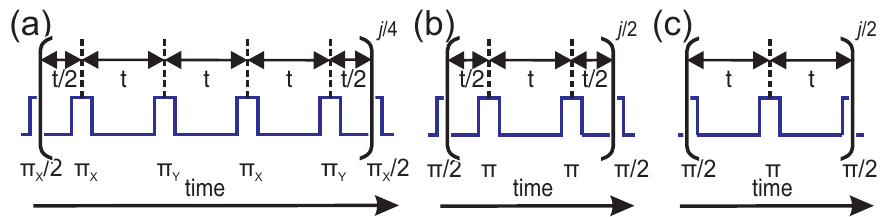}%
\end{center}
\caption{Schematic Sequences, (a) XY4-$N$ sequence with $N$ being the total number of $\pi$-pulses. As the rotation axes $\uvect x$ and $\uvect y$ have the same effect on the time evolution operator, it can be deduced that the periodicity is shorter as indicated in (b). (c) shows the PDD sequence.}
\label{fig_scheme_seq}%
\end{figure}

Because of the large zero-field splitting of the NV-spin $\vect S$ and moderate external magnetic fields, its interaction with the nearby nuclear spin $\vect I$ is treated within the secular approximation ($\vect S = S_z \uvect z$). The effect of the external magnetic field on the electron is undone by the echo sequence and therefore neglected ($\hbar = 1$). The reduced Hamiltonian is
\begin{equation}
H = S_z \vect A \cdot \vect I + \mu_{\rm n} \vect I \cdot \vect B
\label{eq_Hamiltonian}
\end{equation}
As the Hamiltonian is diagonal with respect to the NV spin, the nuclear spin evolves conditionally on the NV spin state in an effective magnetic field $\vect B_{m_s} = \omega_{m_s} \uvect n_{m_s}$; ($\omega_{m_s} = \left| \mu_n \vect B+ m_s \vect A \right|, \uvect n_{m_s}= (\mu_n \vect B+ m_s \vect A )/\omega_{m_s} $).
\begin{equation}
H_{\rm nuc}[m_s] = \omega_{m_s}/2 \boldsymbol \sigma \cdot \uvect n_{m_s}
\label{eq_nuc_H}
\end{equation} 
In the experiment, the NV spin states $m_s = \pm 1$ are detuned, we can therefore treat the NV spin as a two-level system; here we restrict ourselves to $m_s =0$ and $m_s =1$.
In the applied echo sequence the NV spin state is initially prepared in the $+\sigma_x$ direction (in the rotating frame); the nuclear state is in a completely mixed state, the initial state of the system is therefore given by the density matrix.
\begin{equation}
\rho_0 = \tfrac{1}{2}(\mathbb{1}_S+\sigma_x) \otimes \tfrac{1}{2} \mathbb{1}_I
\label{eq_densityMatrix}
\end{equation}
As the final $\pi$/2-pulse rotates about the same axis as the initial one, we measure the expectation value $\braket{\sigma_x}$ (referred to as $\mathcal S$ in the main text). The fluorescence rate (experimental signal $p$) is simply derived from this as
\begin{equation}
p = \frac{1}{2}(1+\braket{\sigma_x})
\label{eq_sig_def}
\end{equation}
The expectation value is rewritten as ($\Re$ designates the real part)
\begin{equation}
\braket{\sigma_x} \equiv \Tr[\rho \sigma_x] 
= \Re \Tr[\rho(t) \sigma_+] 
= 2\Re \Tr_\textrm{nuc}\braket{0|\rho|1}
\label{eq_signalDef}
\end{equation}
This is structurally similar to an interference amplitude of the two NV spin states $\ket 0, \ket 1$, subject to some evolution described by $\rho(t)$. 
Throughout the sequence, the NV spin is flipped $m_s =0 \leftrightarrow m_s=1$; in total $N$ times, $N$ is even (odd) in case of XY4 (PDD). Figure\,\ref{fig_scheme_seq} schematically shows the microwave manipulation of the sequences.
Depending on the sequence the rotation axes are either $\uvect x$ (PDD and CP) or alternating $\uvect x$ and $\uvect y$ (XY4). 
%
%To simplify notation in the following equation we set $H_{\rm nuc}[-1] \equiv H_{\rm nuc}[0]$; the time evolution $mathcal U$ takes the form
The time evolution is generally given by
\begin{align}
\mathcal U  =& \ldots \exp[-\tfrac{\mi}{\hbar} H_{\rm nuc}[m_s]\tau_2] \exp[-\mi \pi/2 \sigma_x] \exp[-\tfrac{\mi}{\hbar} H_{\rm nuc}[m_s]\tau_1]  
%\nonumber \\
%=& \ldots  \exp[-\tfrac{\mi}{\hbar} H_{\rm nuc}[m_s]\tau_2]  \exp[-\tfrac{\mi}{\hbar} H_{\rm nuc}[-m_s]\tau_1] \exp[-\mi \pi/2 \sigma_x]
\label{eq_timeEvol}
\end{align}
Inserting \eqref{eq_densityMatrix},\eqref{eq_timeEvol} in \eqref{eq_signalDef} and utilizing the commutator relations 
\begin{align}
& \exp[\mi \pi/2 \sigma_{x,y}] H_{\rm nuc}[0] = H_{\rm nuc}[1] \exp[\mi \pi/2 \sigma_{x,y}]~\textrm{and vice versa} \nonumber\\
& \exp[\mi \pi/2 \sigma_{y}] \sigma_x = -\sigma_x \exp[\mi \pi/2 \sigma_{y}]
\label{eq_comm_rel}
\end{align}
we obtain
\begin{align}
\braket {\sigma_x}= \tfrac{1}{2} \Re \underbrace{\braket{0| (\mathbb 1_S+ (-1)^{j}\sigma_x)|1}}_{(-1)^{j}}
\Tr_\text{nuc} [\exp[\mi H_{\rm nuc}[0] \tau_n ] \exp[\mi H_{\rm nuc}[1] \tau_{n-1} ]
[\ldots] \exp[-\mi H_{\rm nuc}[0] \tau_{n-1}] \exp[-\mi H_{\rm nuc}[1] \tau_n] ],
%_{\equiv \mathcal U^\dagger_1[\tau_{n-1}]}
\label{eq_unitary_evol}
\end{align}
where $j$ designates the total number of applied $\pi$-pulses around the y-axis. This is the only difference between a Carl-Purcell (CP-$N$) and XY4-$N$ sequence\,\cite{Ryan2010}; in accordance with most publications we ignore that sign and set it to +1. 
%Here the time evolution operator for the nuclear spin $\mathcal U_{m_s}[\tau]$ is introduced.
The expression will now be evaluated in the case of PDD-$N$ and (XY4)-$N$ sequences. 
We identify repeating parts of the sequence and describe their total evolution by a single effective rotation around $\uvect n$ with angle $\phi$, both depending on the evolution time $t$. From the last equation it is clear that a XY4-$N$ sequence can be split into smaller repeating parts (referred to as XY below) (see fig.\,\ref{fig_scheme_seq}b). In the experiment the XY4 sequence is utilized, as it decreases the impact of pulse imperfections compared to XY or PDD.

The repetition of these parts can then conveniently be carried by multiplying the total angle by the number of repetitions $k$. In case of XY4 (PDD) it applies $k = N/2$ ($k = (N+1)/2$)
\begin{align}
\mathcal U_\uvect n \equiv (\mathcal U_\textrm{rep})^k = (\mathbb 1 \cos \phi + \mi \boldsymbol \sigma \underbrace{\uvect n \sin \phi}_{\equiv \vect n} )^k = \mathbb 1 \cos k\phi + \mi \boldsymbol \sigma \uvect n \sin k\phi %= \mathbb 1 T_k(\cos \phi)+\mi \sigma \uvect n \sin \phi U_{k-1}(\cos \phi)
%\label{eq:}
\end{align}
In equation~\eqref{eq_unitary_evol} the product of two such rotations is regarded, arising from the $\ket 0$ and $\ket 1$ state of the NV spin. With the assumption that both rotations employ the same effective angle $\phi$, yet different rotation axes $\uvect n, \uvect m$ we obtain (in the last step the Chebyshev polynomials are utilized).
\begin{align}
\mathcal U_\uvect n \mathcal U_\uvect m = & \mathbb 1 (\cos^2 k \phi + \uvect n \uvect m \sin^2 k \phi)
+ \mi \boldsymbol \sigma ((\uvect n + \uvect m) \sin k \phi \cos k \phi - \uvect n \times \uvect m \sin^2 k \phi) \nonumber \\
=& \mathbb 1(T_k^2(\cos \phi) + \vect n \vect m U_{k-1}^2(\cos \phi) ) + \mi \boldsymbol \sigma
((\vect n + \vect m) T_k(\cos \phi) U_{k-1}(\cos \phi) - \vect n \times \vect m U_{k-1}^2(\cos \phi))
%\label{eq:}
\end{align}
Inserting this into eq.~\eqref{eq_unitary_evol} we obtain (utilizing Pell's equation $T_k^2(\cos \phi)+\cos^2\phi U_{k-1}^2(\cos \phi)=1$)
\begin{align}
\label{eq_gen_result}
\braket{\sigma_x} =& T_k^2(\cos \phi) - \vect n \vect m U_{k-1}^2(\cos \phi)
=1-(1+\vect n \vect m - \cos^2 \phi)U_{k-1}^2(\cos\phi)\\
=&T_k^2(\cos \phi) - \frac{\vect n \vect m}{1-\cos^2 \phi} (1- T_k^2(\cos \phi))
\label{eq_gen_result_bound}
\end{align}
The second equation is useful to determine the boundaries of the signal for all possible number of repetitions $k$.

\subsection{XY4 Sequence}

\begin{figure}[!ht]
\begin{center}
\includegraphics[scale=.75]{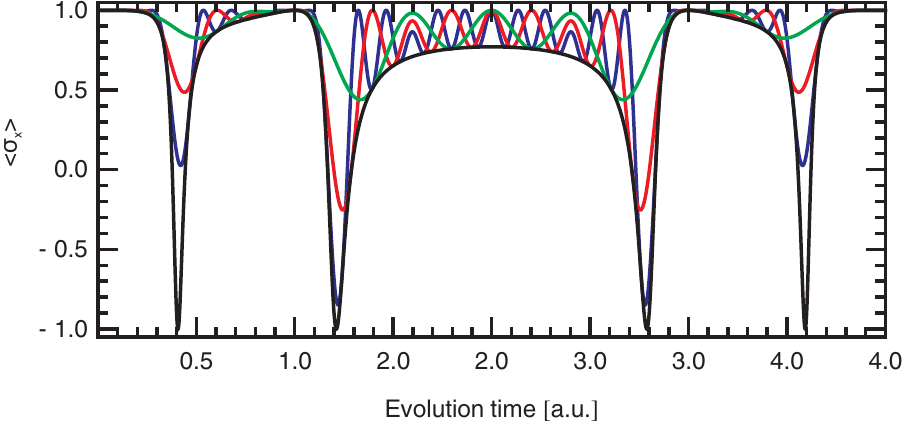}%
\end{center}
\caption{ Signal strengths for varying number of evolutions (green, red, blue: XY 2,4,6; respectively); the black curve is the lower boundary to all the possible XY sequences.} 
\label{fig_xy4}%
\end{figure}

In the case of XY4 the iterative part of the sequence can be expressed as (see fig.~\ref{fig_scheme_seq}); we use the abbreviation $\phi_{m_s} = \omega_{m_s} t/2$
\begin{align}
\mathcal U_1[t/2] \mathcal U_0[t] \mathcal U_1[t/2] 
 = &\mathbb{1} (\cos\phi_0 \cos\phi_1- \uvect n_0 \cdot \uvect n_1 \sin \phi_0 \sin\phi_1 ) \\
+ &\mi \boldsymbol{\sigma}  \cdot (- \uvect n_0 \sin\phi_0 \cos \phi_1 - \uvect n_1 \sin \phi_1 \cos\phi_0 + 2 \uvect n_1 \times (\uvect n_1 \times \uvect n_0 ) \sin^2 \phi_1/2 \sin \phi_0)\nonumber \\
\equiv & \mathbb 1 \cos \phi + \mi \boldsymbol \sigma \uvect n \sin \phi 
= \mathbb 1 \cos \phi + \mi \boldsymbol \sigma \vect n \nonumber
%\label{eq:}
\end{align}
and
\begin{align}
\label{eq_XYsig1}
\mathcal U_0^\dagger [t/2] \mathcal U_1^\dagger[t] \mathcal U_0^\dagger [t/2] 
 = &\mathbb{1} (\cos\phi_0 \cos\phi_1- \uvect n_0 \cdot \uvect n_1 \sin \phi_0 \sin \phi_1 ) \\
+ &\mi \boldsymbol{\sigma}  \cdot ( \uvect n_0 \sin\phi_0 \cos \phi_1 + \uvect n_1 \sin \phi_1 \cos\phi_0 - 2 \uvect n_0 \times (\uvect n_0 \times \uvect n_1 ) \sin^2 \phi_0/2 \sin \phi_1)\nonumber \\
\equiv & \mathbb 1 \cos \phi + \mi \boldsymbol \sigma \uvect m \sin \phi \nonumber
\end{align}
Inserting $\vect n, \vect m$ and $\cos \phi$ into eq.~\eqref{eq_gen_result}, we obtain the explicit result with $\cos \phi$ given by the preceding equations.
\begin{align}
\label{eq_XYsig2}
\braket {\sigma_x}
= 1-4 (1-(\uvect n_0 \uvect n_1)^2) \sin^2 \phi_0/2 \sin^2 \phi_1/2 \left(1-\cos \phi \right) U^2_{k-1}(\cos \phi)
\end{align}
We now regard the solution written in the form \eqref{eq_gen_result_bound}.
As $0 \leq\ T^2_k[\left| x\right|<1] \leq 1$, the fraction in the equation represents the lower limit of the signal independent of number of repetitions $k$, see fig.\ref{fig_xy4}. In particular, the absolute minimum is obtained when the axes of the effective rotations $\vect n, \vect m$ are parallel. As they have the same length, they are identical $\vect n = \vect m$. Inserting this into eqs.~\eqref{eq_XYsig1},\eqref{eq_XYsig2} yields the two conditions.
\begin{align}
 &\sin \phi_0 \cos \phi_1 + \sin^2 \phi_1/2 \sin \phi_0 - \uvect n_0 \uvect n_1 \sin^2 \phi_0/2 \sin \phi_1 =0  \\
&\sin \phi_1 \cos \phi_0 + \sin^2 \phi_0/2 \sin \phi_1 - \uvect n_0 \uvect n_1 \sin^2 \phi_1/2 \sin \phi_0 =0 \nonumber
%\label{eq:}
\end{align}

\subsection{PDD Sequence}

The evolution can be written as
\begin{align}
\mathcal U_0[t] \mathcal U_1[t]  
 = &\mathbb{1} (\cos\phi_0 \cos\phi_1- \uvect n_0 \cdot \uvect n_1 \sin \phi_0 \sin\phi_1 ) \\
+ &\mi \boldsymbol{\sigma}  \cdot (\uvect n_0 \sin \phi_0 \cos \phi_1 + \uvect n_1 \sin \phi_1 \cos \phi_0 - \sin \phi_0 \sin \phi_1 \uvect n_0 \times \uvect n_1) \nonumber \\
\equiv & \mathbb 1 \cos \phi + \mi \boldsymbol \sigma \uvect n \sin \phi 
= \mathbb 1 \cos \phi + \mi \boldsymbol \sigma \vect n \nonumber
%\label{eq:}
\end{align}
$\uvect m$ is obtained by reversing the angles $\phi_0 \rightarrow -\phi_0,\phi_1 \rightarrow -\phi_1$, inserting this into eq.~\eqref{eq_gen_result} yields the result for PDD.
\begin{align}
\label{eq_PDDsig}
\braket {\sigma_x}
= 1-2 (1-(\uvect n_0 \uvect n_1)^2) \sin^2 \phi_0 \sin^2 \phi_1 U^2_{k-1}(\cos \phi)
\end{align}
Which is easily seen to reduce to the well known result for Hahn spin-echo as $U_{k-1=0}=1$.

\section{Background due to the Spin Bath}

We treat the impact of the bath of \C as a classical magnetic field, whose frequency distribution $\bar S(\omega)$ is Gaussian around the Larmor precession due to the external field, giving rise to decaying collapses and revivals. Please note that this does not include the dominant part of the decoherence, which is a result of low-frequency noise due to flip-flop processes within in the bath, see e.g.\,\cite{Sousa2009},\cite{Kolkowitz23022012}. The coherence envelope $\braket{\braket{\sigma_+}}$ in the system may be written as
\begin{equation}
\braket{\braket{\sigma_+}} = \exp [-\chi[t]] \equiv \exp \left[- \int \frac{d\omega}{2\pi}\frac{F(\omega t)}{\omega^2}\bar S(\omega) \right]
\label{eq_filterfunc}
\end{equation}
Here $F(z)$ defines the sequence-dependent filter function; it takes the form 
\begin{equation}
F(z) = a_0  + \sum_n a_n \cos(n z/2)
\label{eq_filterfunc1}
\end{equation}
Assuming the width $\sigma$ of the Gaussian distribution to be narrow $\sigma \ll \omega_0$ we can simplify it as
\begin{equation}
\bar S(\omega) = \tilde \lambda^2 \omega_0^2 \exp[-(\omega-\omega_{\rm L})^2/(2 \sigma^2)] \approx \tfrac{\sqrt{2 \pi}}{\sigma}\lambda^2 \omega^2\exp[-(\omega-\omega_{\rm L})^2/(2 \sigma^2)]
\label{eq_larmor_noise}
\end{equation}
$\chi(t)$ can be easily integrated piecewise, yielding 
\begin{equation}
\chi(t) = \lambda^2 \left(a_0 +\sum_n a_n \exp(\tfrac{1}{8} n^2 t^2 \sigma^2) \cos(n t \omega_0)\right)
\label{eq_gauss_dec}
\end{equation}
\begin{figure}[ht!]
\includegraphics{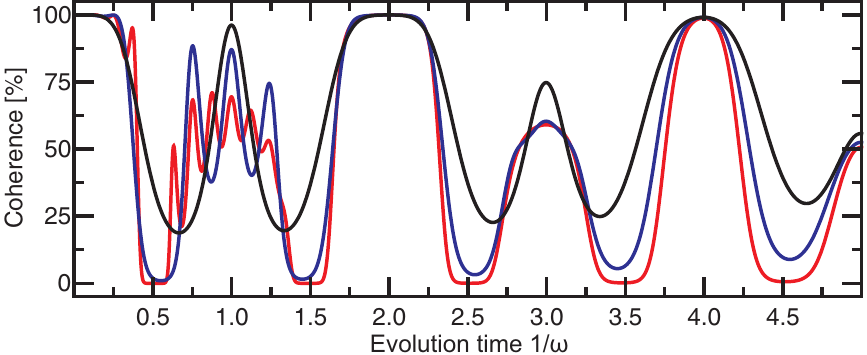}%
\caption{Simulated bath decoherence, XY-2,4,8 (black, blue, red, respectively) $\sigma = 0.1 \omega/(2 \pi), \lambda = 0.25$. Faster oscillations can be seen to be damped more quickly.}%
\label{fig_bg}%
\end{figure}
The high frequency components in the filter function die out faster, as shown in fig.\,\ref{fig_bg}. 

In addition, the low frequency components in magnetic noise lead to a dominant overall decay in coherence. Increasing the number $N$ of $\pi$-pulses in a sequence reduces the susceptibility to low frequency noise and thereby enhances the effective coherence time $T_2^{\rm eff}$, see.\,\cite{Ryan2010},\cite{DeLange2010}.
\begin{equation}
\braket{\braket{\sigma_+}}_{\rm low freq} = \exp\left[-k \left( \tfrac{2 t}{T_2} \right)^3 \right] \equiv \exp \left[- \left( \tfrac{2 t}{T_2^{\rm eff}} \right)^3 \right]
\label{eq_low_freq_noise}
\end{equation}
We include this term when we regard a revival of the bath induced coherence (fig.\,3 in the main part) and use it for scaling arguments when analyzing the limits of our method.

\section{Approximation of the Hyperfine Interaction}

\begin{figure}[ht!]
\includegraphics{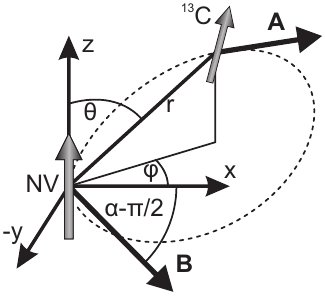}%
\caption{Definition of the coordinate system. The $z$-axis is defined by the symmetry axis of the NV center, while the $x$-axis is defined such that the magnetic field vector $\vect B$ always lies within the $x$-$z$ plane as the magnet is rotated. The angle of magnetic field rotation $\alpha$ is defined relative to the $z$-axis, but the arc is shown here without the first  $\pi/2$ segment from the $z$ to $x$ axes for clarity. The position of the \C relative to the NV is then given by the spherical coordinates $r$, $\phi$, and $\theta$, while $\vect A$ is the hyperfine coupling vector resulting from this spatial configuration.}%
\label{fig_coord_sys}%
\end{figure}

For the following discussion, we regard the situation in which the coordinate system is chosen such that the $z$-axis coincides with the symmetry axis of the NV center, the NV spin's center of mass defines the origin, and the \C lies within the $x$-$z$ plane, (a special case of the more general case shown in fig.\,\ref{fig_coord_sys}). Motivated by its small magnitude, we treat the hyperfine coupling as rotationally symmetric about $z$, reducing the hyperfine interaction $\vect A$ to $\vect A = \{A_x, 0, A_z\}$. Note that the contact interaction is still treated exactly within this simplification. As the NV wavefunction is not perfectly rotationally symmetric, it can potentially produce a magnetic field at the \C location that has components in the $\uvect y$ direction, which we neglect here. From the consistency of our fits we infer that these terms are small. 
	
	Previous works have observed an enhanced \C gyromagnetic ratio for transverse fields, arising from coupling to the NV electronic spin (see for example \cite{Childress2006}.) This enhancement of the gyromagnetic ratio is a result of corrections to the secular approximation that we do not consider as they are negligible when the coupling $\vect A$ is weak enough such that $\mu_e/\mu_n \ll \Delta/\left| \vect A \right|$, as is the case for all \C spins considered in this work.

\section{Signal Strength vs. Number of Applied Pulses}

When the hyperfine coupling strength $\left| \vect A \right|$ exceeds the bare Hahn echo decoherence rate $1/T_2$, a strong signal can be obtained without enhancing the coherence to $T_2^{\rm eff}$ using extended sequences. We therefore numerically investigate the maximal impact $p_{\rm max}$ of an individual \C in the regime $1/T_2>\left| \vect A \right|$ using extended sequences.

\begin{figure}[ht!]
\begin{center}
\includegraphics{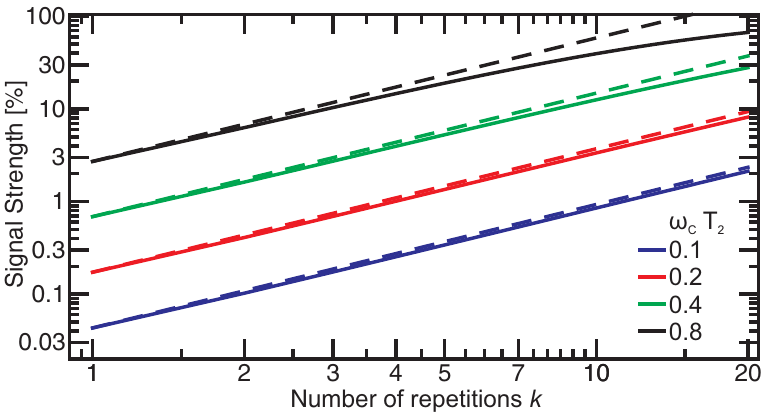}%
\end{center}
\caption{Numerically calculated signal strength vs. number of pulses and coherence time; the dashed lines highlight the scaling as $k^{4/3}$} 
\label{fig_signal_strength}%
\end{figure}

At first we will consider the impact arising form the coherent evolution described by eq.\,\eqref{eq_XYsig2} which is suppressed by low frequency noise, see eq.\,\eqref{eq_low_freq_noise}. We therefore regard an expression of the form
\begin{equation}
p = \tfrac{1}{2}\left[1-\braket{\sigma_x} \right]\braket{\braket{\sigma_+}}
\label{eq_signal_strength}
\end{equation}
We calculate the maximum signal $p_{\rm max}$ for a given coupling strength $\left| \vect A \right|$, the orientation and strength of the external magnetic field as well as the evolution time $t$ are optimized numerically. Figure\,\ref{fig_signal_strength} shows the result.
The signal strength can be seen to scale with the number of repetitions $k$ as $k^{4/3}$; this can be understood as follows: ideally, the two states of the \C superposition get pushed apart linearly with the number of repetitions $k$ and the evolution time $t$. For small differences in these states, this leads to a quadratic increase in signal strength.
One therefore obtains
\begin{equation}
p \propto t^2 k^2 \exp \left[-k \left( \tfrac{2 t}{T_2} \right)^3 \right];~~
\frac{\partial p}{\partial t} = 0 ~~ \Rightarrow ~~ t \propto k^{-1/3}~ {\rm and}~ p_{\rm max} \propto k^{4/3}
\label{eq_signal_scaling1}
\end{equation}
If the investigated spin is a \C, one has to include the collapses and revivals in coherence coming from the Larmor precession (at frequency $\omega_{\rm L}$ of the \C bath spins.
For $\omega_{\rm L} \gg \left| \vect A \right|$ the signal $p$ tends to peak at $t \approx \pi/\omega_{\rm L}(1/2+n),~n$: integer; however the signal in this range is exponentially suppressed due to the collapses from the bath. 

%Additionally, another difficulty arises when searching for \Cs, as the signal strength scales less favorable with non optimized parameters.

\section{Ramsey Measurements}

\begin{figure}[!htb]
\begin{center}
\includegraphics{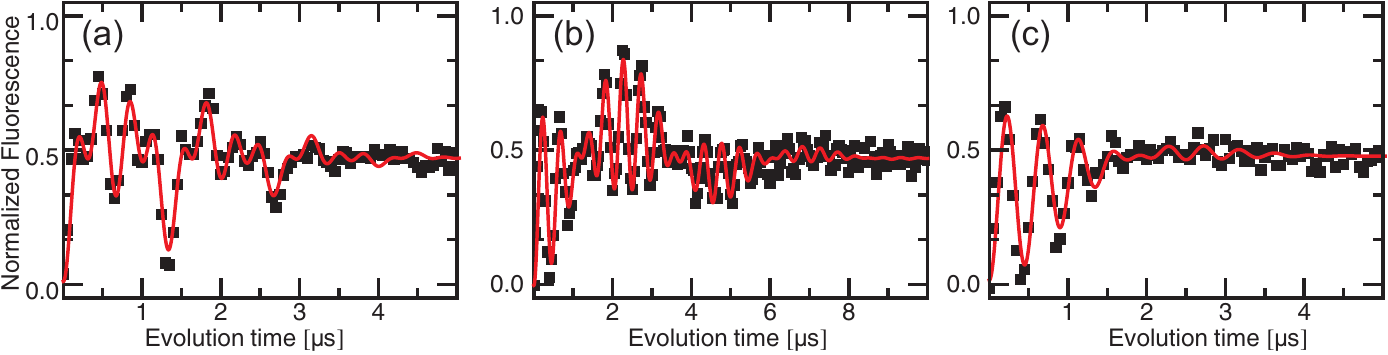}%
\end{center}
\caption{Ramsey measurements of the NV centers investigated in the main text. Data shown in (a), (b) and (c), corresponds to the NV data shown in fig.\,2,\,3\,4 in the main text. Extracted $T_2^\ast$ are $2.5\pm 0.1,4.6\pm 0.2,1.8\pm0.2$\,\textmu s, respectively.}
\label{fig_ramsey}% 
\end{figure}

We infer the inhomogeneous dephasing rate $1/T_2^\ast$ of the NV spin from Ramsey measurements, as shown in fig.\,\ref{fig_ramsey}. The result can be accurately reproduced taking only the coupling to the nuclear spin of the nitrogen into account (coupling strength: 2.2\,MHz). We measure up to an evolution time of 40\,\textmu s to ensure that there are no revivals in coherence. 
%In accordance with our expectation, the NV with the weakest coupled \C has the smallest inhomogeneous coherence rate.
%

\section{Consistency of Ramsey and XY4 Measurements}

\begin{figure}[!htp]
\begin{center}
\includegraphics{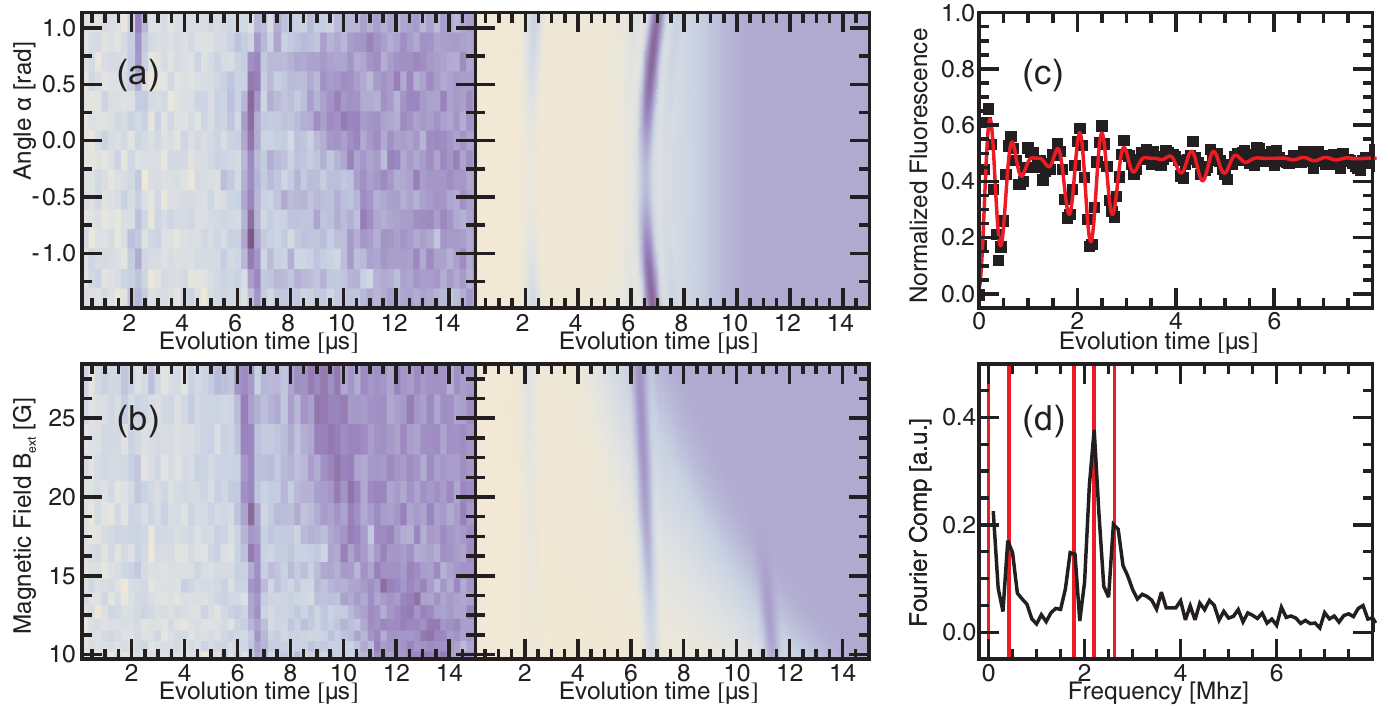}%
\end{center}
\caption{XY4-8 and Ramsey measurements. (a) and (b) display XY4-8 measurement and fit vs. magnetic field orientation and strength, respectively. (c) shows a Ramsey measurement performed on the same NV, it is fitted incorporating an additional energy splitting due to the coupling to the \C. (d) is the fourier transform of the data shown in $($c$)$. In addition to the 2.2 MHz splitting from the nitrogen, a clear splitting due to the hyperfine coupling is visible; lines mark the position of the peaks as determined from (c).}
\label{fig_ramsey_XY4}% 
\end{figure}

As a consistency check of our method, we show in fig.\,\ref{fig_ramsey_XY4} measurements performed on an NV center with a strongly coupled \C, where $\vect A>1/T_2^\ast$. We measure its impact on the NV spin evolution in a XY4-8 measurement (a,b) as well as in a Ramsey sequence (c,d). The extracted hyperfine couplings are 422\,kHz (Ramsey) and 438\,kHz (XY4-8); these values are larger the inhomogeneous dephasing rate 286\,kHz.

\section{Additional XY4-8 Measurements - Complementing Figs. 2,4}

\begin{figure}[!htp]
\begin{center}
\includegraphics{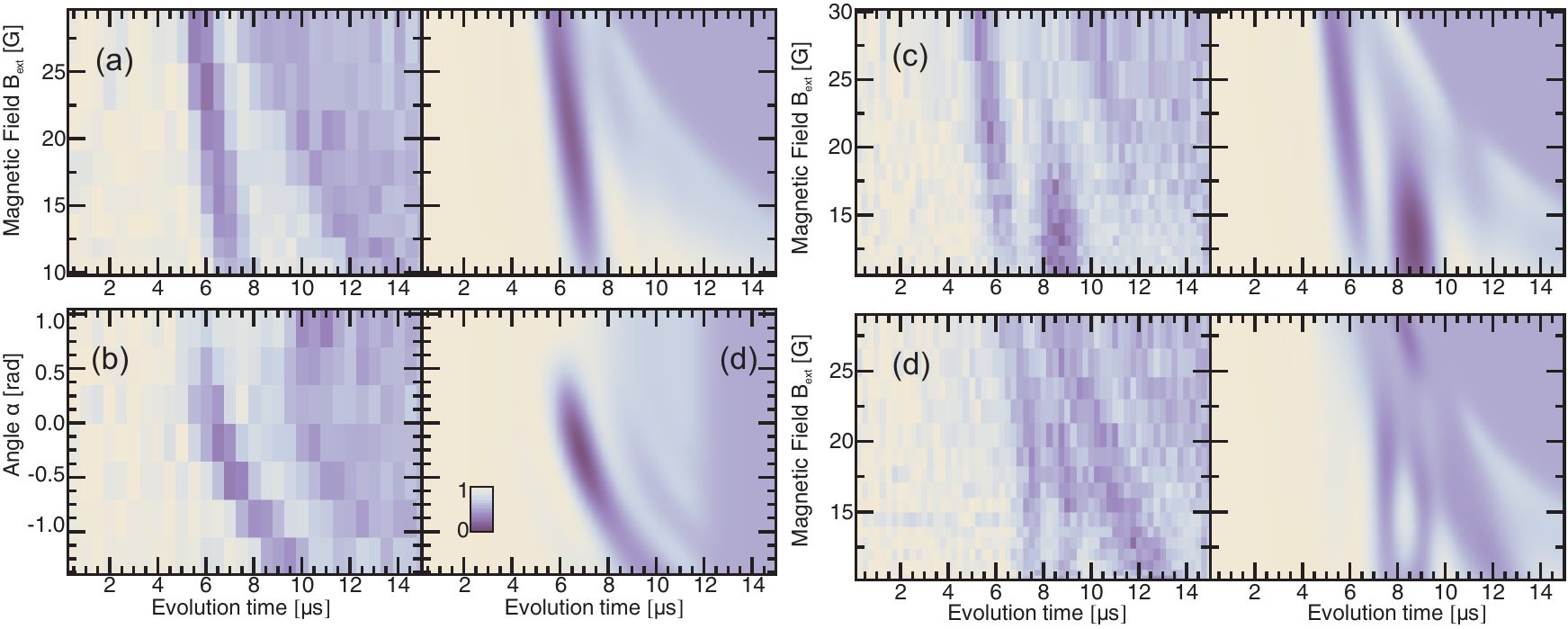}%
\end{center}
\caption{Measurements and simulation completing the data sets shown in the main text, showing the NV response vs magnetic field strength (a,c,d) and orientation (b). (a and b) Signal when addressing the lower transition of the NV shown in fig.\,2 (main text). (c,d) The data set corresponding to the NV shown in fig.\,4 of the main text; (c): upper transition, (d): lower transition.
}
\label{fig_XY4}% 
\end{figure}

In fig.\,\ref{fig_XY4} we show the remaining data from the data sets used to generate the fits shown in figs.\,2 and 4 of the main text. In general, we measure the NV response against magnetic field strength and orientation for both the upper and lower NV spin transitions.

\section{Additional XY4 Measurements - Complementing Fig. 3}

\begin{figure}[!htp]
\begin{center}
\includegraphics{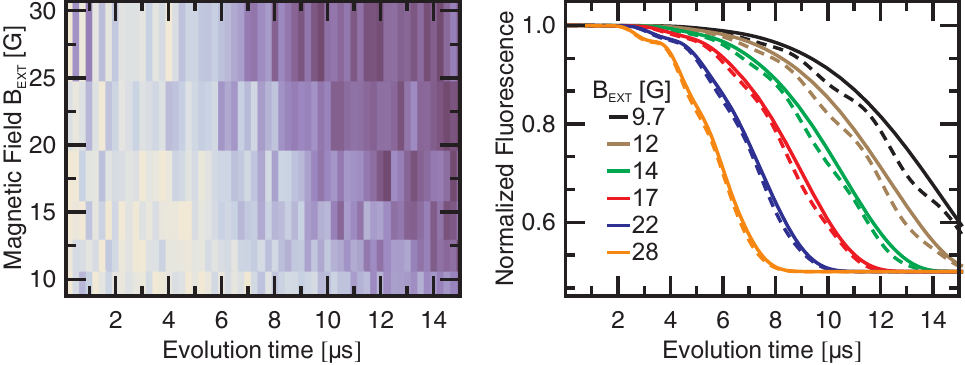}%
\end{center}
\caption{(a) Measurement of NV subject to a XY4-16 sequence. (b) Calculated response (solid lines) of the NV obtained by fitting~\eqref{eq_gauss_dec} to the data shown in (a); dashed lines include the additional effect of the isolated nearby \C based on the coupling parameters extracted from the fit to the data shown in fig.\,3 in the main text.}
\label{supp_fig3}% 
\end{figure}

In fig.\,\ref{supp_fig3} we show additional data for the NV with the weakest coupled \C observed, (fig.\,3 of the main text). The experimental settings are similar to those of figs.\,2 and 4 in the main text, but XY4-16 is used in place of XY4-8 to further enhance any impact of the weakly coupled \C. No clear impact from the \C is observed, which is consistent with our calculated response based on the fit displayed in the main text.

\section{Parameter table of investigated NV Centers}

In table\,\ref{tab_param} we present the extracted parameters from the NV centers we observed that could be treated within the framework derived above. In particular, NVs with strongly coupled \Cs ($\left| \vect A \right|>$ 1MHz) were not investigated in detail. The hyperfine coupling strengths $\vect A$ are determined from the fits to the XY-4 measurements; the extracted distance $r$ and angle $\theta$ (defined in fig.\,\ref{fig_coord_sys},) are based on the assumption of point-like dipole interactions, see main text for details and discussion.

\begin{table}[!htp]
\begin{center}
\begin{tabular}{ccccccccc}
NV index (arb.) & $T_2^\ast$ [\textmu s] &  number of \Cs  & $A_x$ [kHz] & $A_z$ [kHz] &  $\left| \vect A \right|$ [kHz] &  distance $r$  [\AA] & angle $\theta$ [rad] & appears in fi	g.\\
\hline
1		& 2.5 & 1 & 100 & -75 & 126	& 5.6 & 18 	& 2,\ref{fig_ramsey},\ref{fig_XY4} \\

2 	& 4.6 & 1	& -40	& -22 &	46	& 8.0 & -22 & 3,\ref{fig_ramsey} \\

3		& 3.5 & 1 & -222& -359& 422 & 3.6 & -10 & \ref{fig_ramsey_XY4} \\

4		& 2.0 & 2	& 180	&	-51	& 187	&	5.1 & 27	& - \\
		& 		&   & 63 	& -61 & 88	& 6.3 & 15	& - \\

5		& 1.8 & 3 & 98	&	97	&	138	&	6.3 & 60	& 4,\ref{fig_ramsey},\ref{fig_XY4} \\
		& 		&  	& 115	& -13	&	116	& 6.1	& 32	& 4,\ref{fig_ramsey},\ref{fig_XY4} \\
		& 		&   & 37	& -52 & 64	& 6.9 & 12	& 4,\ref{fig_ramsey},\ref{fig_XY4} \\
						
\end{tabular}
\end{center}
\caption{Extracted parameters from XY4-8 and Ramsey measurements.}
\label{tab_param}
\end{table}